\documentclass[12pt]{article}
\usepackage{amsfonts,amsmath,amssymb,mathrsfs,natbib,url}

\title{Paradoxes and Primitive Ontology in\\ Collapse Theories of Quantum Mechanics}
   
\author{
Roderich Tumulka\footnote{Department of Mathematics,
     Rutgers University, Hill Center,  
     110 Frelinghuysen Road, Piscataway, NJ 08854-8019, USA.
     E-mail: tumulka@math.rutgers.edu}
}

\date{February 28, 2017}

\newcommand{\ket}[1]{|#1\rangle}
\newcommand{\pr}[1]{|#1\rangle\langle #1|}
\newcommand{\dead}{\ket{\text{dead}}}
\newcommand{\alive}{\ket{\text{alive}}}
\newcommand{\scp}[2]{\langle #1|#2 \rangle}
\newcommand{\be}{\begin{equation}}
\newcommand{\ee}{\end{equation}}
\newcommand{\tr}{\mathrm{tr}}
\newcommand{\vQ}{\boldsymbol{Q}}

\begin{document}
\maketitle
\begin{abstract}
Collapse theories are versions of quantum mechanics according to which the collapse of the wave function is a real physical process. They propose precise mathematical laws to govern this process and to replace the vague conventional prescription that a collapse occurs whenever an ``observer'' makes a ``measurement.'' The ``primitive ontology'' of a theory (more or less what Bell called the ``local beables'') are the variables in the theory that represent matter in space-time. There is no consensus about whether collapse theories need to introduce a primitive ontology as part of their definition. I make some remarks on this question and point out that certain paradoxes about collapse theories are absent if a primitive ontology is introduced.

\medskip

\end{abstract}

\section{Introduction}

Although collapse theories \citep{Ghi07} have been invented to overcome the paradoxes of orthodox quantum mechanics, several authors have set up similar paradoxes in collapse theories. I argue here, following \citet{Mon04}, that these paradoxes evaporate as soon as a clear choice of the primitive ontology is introduced, such as the flash ontology or the matter density ontology. In addition, I give a broader discussion of the concept of primitive ontology, what it means and what it is good for.

According to collapse theories of quantum mechanics, such as the Ghirardi--Rimini--Weber (GRW) theory \citep{GRW86,Bell87} or similar ones \citep{Pe89,Dio89,BG03}, the time evolution of the wave function $\psi$ in our world is not unitary but instead stochastic and non-linear; and the Schr\"odinger equation is merely an approximation, valid for systems of few particles but not for macroscopic systems, i.e., systems with (say) $10^{23}$ or more particles. The time evolution law for $\psi$ provided by the GRW theory is formulated mathematically as a stochastic process, see, e.g., \citep{Bell87,BG03,AGTZ06}, and can be summarized by saying that the wave function $\psi$ of all the $N$ particles in the universe evolves as if somebody outside the universe made, at random times with rate $N\lambda$, an unsharp quantum measurement of the position observable of a randomly chosen particle. ``Rate $N\lambda$'' means that the probability of an event in time $dt$ is equal to $N\lambda \, dt$; $\lambda$ is a constant of order $10^{-15}\, \mathrm{sec}^{-1}$. It turns out that the empirical predictions of the GRW theory agree with the rules of standard quantum mechanics up to deviations that are so small that they cannot be detected with current technology \citep{BG03,Adl07,FT12,BU14,CBFV16}.

The merit of collapse theories, also known as dynamical state reduction theories, is that they are ``quantum theories without observers'' \citep{Gol98}, as they can be formulated in a precise way without reference to ``observers'' or ``measurements,'' although any such theory had been declared impossible by Bohr, Heisenberg, and others. Collapse theories are not afflicted with the vagueness, imprecision, and lack of clarity of ordinary, orthodox quantum mechanics (OQM). Apart from the seminal contributions by \citet{GRW86,Bell87,Pe89,Dio89,Dio90}, and a precursor by \citet{Gis84}, collapse theories have also been considered by \citet{GP93,Leg02,Pen00,Adl07,Wein12}, among others. A feature that makes collapse models particularly interesting is that they possess extensions to relativistic space-time that (unlike Bohmian mechanics) do not require a preferred foliation of space-time into spacelike hypersurfaces \citep{Tum06,Tum06c,BDGGTZ}; see \citet{Mau11} for a discussion of this aspect.

Collapse theories have been understood in two very different ways: some authors [e.g., \citet{Bell87,BGG95,Gol98,Mau06,AGTZ06,Esf14}] think that a complete specification of a collapse theory requires, besides the evolution law for $\psi$, a specification of variables describing the distribution of matter in space and time (called the \emph{primitive ontology} or PO), while other authors [e.g., \citet{AL90,Shi,Lew95,Pen00,Adl07,Pea06,Alb15}] think that a further postulate about the PO is unnecessary for collapse theories. The goals of this paper are to discuss some aspects of these two views, to illustrate the concept of PO, and to convey something about its meaning and relevance. I begin by explaining some more what is meant by ontology (Section~\ref{sec:ontology}) and primitive ontology (Section~\ref{sec:PO}). Then (Section~\ref{sec:paradoxes}), I discuss three paradoxes about GRW from the point of view of PO. In Section~\ref{sec:need}, I turn to a broader discussion of PO. Finally in Section~\ref{sec:mind}, I describe specifically its relation to the mind-body problem.

\section{Ontology and Its Relevance to Physics}
\label{sec:ontology}

The ``ontology'' of a theory means what exists, according to that theory. John S.~Bell (\citeyear{Bell76}) coined the word ``beables'' (as opposed to observables) for the variables representing something real, according to that theory; that is, for the ontology. Many researchers, physicists as well as philosophers, find it hard to think in terms of an ontology, partly because we all have practiced for many years to think ``in the quantum mechanical way.'' Physicist Jeremy Bernstein (\citeyear{Ber13}) wrote: 
\begin{quotation}
Many of the papers I tried to read were written by philosophers and had words like ``ontology" sprinkled over them like paprika.
\end{quotation}
The quote conveys hesitation and reservations towards this word. I recommend that physicists overcome this hesitation, as ontology is, in fact, a highly relevant concept. 

And a very simple one. Despite the fancy name, the concept of ontology is already familiar from classical physics. For example, Newtonian mechanics talks about particles, and macroscopic objects such as rocks and trees consist, according to Newtonian mechanics, of particles---so one says that the ontology of Newtonian mechanics is particles. For another example, the electric field may at first appear as a mere calculational device (if defined as the force that a unit charge would feel if placed at a certain location), but later developments (specifically the realization that the electric and magnetic fields can carry energy and momentum) convince us that the electric and magnetic fields are something real. That is, the electric and magnetic fields are part of the ontology of classical electrodynamics.

It was long thought that the key to clarity in QM was to avoid talking about ontology and stick to operational statements. In my opinion, that thought has not paid off. Sometimes, we want to talk about events that occurred before humans existed, and then it seems particularly absurd to assume that facts only become definite when observed. But even when we talk about laboratory experiments, it is hard to stick to operational statements and hard to avoid talking about what actually happened in reality. And about that, OQM has only contradictory and unclear statements to offer. The true key to clarity is to set up a hypothesis about what happens in reality, to analyze what the consequences will be if that hypothesis is taken seriously, and to reject the hypothesis if it leads to empirically incorrect consequences. It was long thought that such hypotheses do not exist for QM, but Bohmian mechanics and collapse theories are counter-examples to that thought. Bohmian mechanics nicely illustrates some advantages of a clear ontology: 
\begin{itemize}
\item The theory can be stated in full on a single page (whereas operational statements are cumbersome to formulate with high precision).
\item The usual quantum rules about the probabilities of outcomes of experiments can be derived as theorems instead of being postulated as axioms \citep{DGZ04}.
\item Paradoxes can be resolved simply by taking seriously the fundamental laws of the theory. For example, \citet{Bell80} analyzed how Wheeler's delayed-choice paradox gets resolved in Bohmian mechanics.
\item Symmetries (such as Lorentz or Galilean invariance or time reversal invariance or gauge invariance) have a clear meaning and can be derived as theorems \citep{AGTZ06}. Likewise for superselection rules \citep{CDT05}.
\end{itemize}

\section{Primitive Ontology in Collapse Theories}
\label{sec:PO}

Primitive ontology (PO) is a name for that part of the ontology that represents matter in $3+1$-dimensional space-time. For example, the PO in classical mechanics and Bohmian mechanics are the particles (material points), mathematically represented by their world lines and thus by their coordinate functions $\vQ_k(t)$ (the position of particle $k$ at time $t$). The wave function in Bohmian mechanics is part of the ontology (it is something real) but not part of the primitive ontology (as it lives in configuration space, not in physical space). 

\citet{Bell76,Bell87} used the expression ``local beables'' for the variables representing something real (``beables'') associated with a space-time point (``local''). For example, in Bohmian mechanics, $\psi$ is non-local because it refers to several space-time points at once, whereas $\vQ_k(t)$ (or, equivalently, the number of actual particles at the space-time point $(x,y,z,t)$) is a local beable. In all theories that I will consider here, the local beables are exactly the primitive ontology (even though it may not necessarily be so in all conceivable theories).

In collapse theories, I will focus on two particular choices of PO for the GRW theory, the flash ontology and the matter density ontology. According to the \emph{flash ontology} \citep{Bell87}, matter is fundamentally described by a discrete set of space-time points called flashes. Flashes are material space-time points. In the GRW theory with the flash ontology (GRWf for short), there is one flash for each collapse of $\psi$; its time is the time of the collapse, and its position is the collapse center; see, e.g., \citep{Bell87,AGTZ06} for mathematical descriptions. 

According to the \emph{matter density ontology} \citep{BGG95,Gol98}, matter is fundamentally continuously distributed and described by a density function $m(x,y,z,t)$ on space-time. In the GRW theory with the matter density ontology (GRWm for short), the $m$ function at time $t$ is obtained from $|\psi_t|^2$ (which is a density function on $3N$-dimensional configuration space) by integrating out the coordinates of $N-1$ particles to get a 3-dimensional density function; in order not to prefer any particle, one averages over all sets of $N-1$ particles (perhaps using a weighted average with the particles' masses as weights). See, e.g., \citep{AGTZ06} for a mathematical description.

The view with which to contrast GRWf and GRWm is that there is no primitive ontology, in fact no ontology at all in physical 3-space, and that the ontology comprises only $\psi$ (GRW$\emptyset$ for short). To illustrate the difference between GRWf/GRWm and GRW$\emptyset$, let me make up a creation myth (as a metaphorical way of speaking): Suppose God wants to create a universe governed by GRW theory. He creates a wave function $\psi$ of the universe that starts out as a particular $\psi_0$ that he chose and evolves stochastically according to a particular version of the GRW time evolution law. According to GRW$\emptyset$, God is now done. According to GRWf and GRWm, however, a second act of creation is necessary, in which he creates the matter, i.e., either the flashes or continously distributed matter with density $m$, in both cases coupled to $\psi$ by the appropriate laws. (I will elaborate on this point in Section~\ref{sec:all}.)

There are several motivations for considering GRW$\emptyset$. First, it seems more parsimonious than GRWm or GRWf. Second, it was part of the motivation behind GRW theory to avoid introducing an ontology in additon to $\psi$; otherwise, we may think, we could have used Bohmian mechanics. In fact, much of the motivation came from the measurement problem of OQM. The measurement problem arises from treating the measurement apparatus as a quantum system (because it consists of electrons and quarks): If the time evolution law for every wave function is linear, then the joint wave function of object and apparatus after a quantum measurement is a non-trivial superposition of states corresponding to different outcomes (except if the object was initially in an eigenstate of the observable measured). If there are no variables in addition to the wave function, then there is no fact about which outcome was the actual one. Thus, if there is an actual outcome then we must either abandon the linearity of the Schr\"odinger evolution or introduce further ontology in addition to $\psi$; this is the measurement problem. The GRW theory was intended to choose the first option, not the second. And yet, I think that GRW$\emptyset$ is not an acceptable theory but GRWm and GRWf are.

In the next section, I will exemplify some issues by means of some paradoxes that have been raised as objections to the GRW theory; I will describe what these paradoxes look like from the point of view of a theory with PO, and will point out some advantages of the PO view by showing that it resolves these paradoxes. In the section after that, I will discuss the need for a primitive ontology more broadly.

\section{Three Paradoxes About GRW Theories}
\label{sec:paradoxes}

In each case, I will first describe a paradox and then explain why it evaporates in GRWf and GRWm. For GRWm, some of the relevant points have already been made by \citet{Mon04}.

\subsection{Paradox 1: Does the Measurement Problem Persist?}

Paradox: Here is a reason why one might think that the GRW theory fails to solve the measurement problem \citep{AL90}. Consider a quantum state like Schr\"odinger's cat, namely a superposition
\begin{equation}\label{cat}
\psi = c_1 \psi_1 + c_2 \psi_2
\end{equation}
of two macroscopically distinct states $\psi_i$ with $\|\psi_1\|=1=\|\psi_2\|$, such that both contributions have nonzero coefficients $c_i$. Given that there is a problem---the measurement problem---in the case in which the coefficients are equal, one should also think that there is a problem in the case in which the coefficients are not exactly equal, but roughly of the same size. One might say that the reason why there is a problem is that, according to quantum mechanics, there is a superposition whereas according to our intuition there should be a definite state. But then it is hard to see how this problem should go away just because $c_2$ is much smaller than $c_1$. How small would $c_2$ have to be for the problem to disappear? No matter if $c_2=c_1$ or $c_2=c_1/100$ or $c_2=10^{-100}c_1$, in each case both contributions are there. But the only relevant effect of the GRW process replacing the unitary evolution, as far as Schr\"odinger's cat is concerned, is to randomly make one of the coefficients much smaller than the other (although it also affects the shape of the suppressed contribution \citep{Wal14}).

\medskip

Answer: The argument indeed raises questions about GRW$\emptyset$. From the point of view of GRWm or GRWf, however, the argument is flawed as it pays no attention to the PO. To take the PO seriously means that whether Schr\"odinger's cat is really dead must be read off from the PO. 

In GRWf, if $|c_2|\ll |c_1|$ then the next flash is overwhelmingly likely to be associated with $\psi_1$ (say, $\psi_1=\dead$), and so on for further flashes, with the result that the flashes form the shape of a dead cat. They form this shape even if there are a few extra flashes associated with $\alive$, but in fact, since each flash associated with a collapse favoring $\dead$ further reduces the probability of a flash associated with $\alive$, it has probability near 1, given that $|c_2|\ll |c_1|$, that the number of later flashes associated with $\psi_2=\alive$ is zero. So yes, the wave function is still a superposition, but the definite facts that our intuition wants can be found in the PO. The flashes represent reality in 4-dimensional space-time; the cat is in the flashes, not in $\psi$. To be sure, we have no precise definition of which patterns of flashes should be regarded as dead cats; that fact, however, is not worrisome; it arises simply because the words ``dead cat'' are not precisely defined in ordinary language.

In GRWm, if $\psi$ is close to $\dead$ then $m$ equals $m_{\dead}$ up to a small perturbation, and that can reasonably be accepted as the PO of a dead cat. Note the following: Whereas the wave function is a superposition of two packets $\psi_1,\psi_2$ that correspond to \emph{two very different} kinds of (particle) configurations in ordinary QM or Bohmian mechanics, there is only \emph{one} configuration of the matter density $m$---the definite fact that our intuition wants. A subtlety here arises from the fact that the small contribution to $m$ that comes from $\psi_2$ often still has internal structure (say, looks like a live cat); I will address this point in Section~\ref{sec:mw} below.

\subsection{Paradox 2: How Can You Call a Cat Dead if There is a Small Probability of Finding it Alive?}

Paradox:  As a variant of the first paradox, one might say that even after the GRW collapses have pushed $|c_1|^2$ near 1 and $|c_2|^2$ near 0 in the state vector \eqref{cat}, there is still a positive probability $|c_2|^2$ that if we make a ``quantum measurement'' of the macro-state---of whether the cat is dead or alive---we will find the state $\psi_2$, even though the GRW state vector has collapsed to a state vector near $\psi_1$, a state vector that might be taken to indicate that the cat is really dead (assuming $\psi_1 = \dead$). Thus, it seems not justified to say that, when $\psi$ is close to $\dead$, the cat is really dead. This paradox is another version of the ``tail problem'' \citep{Shi,AL90,Lew95,AL96,Cor99}.

\medskip

Answer: It is important here to appreciate the following difference between orthodox quantum mechanics and GRWm/GRWf concerning the meaning of the wave function and how the wave function makes contact with our world: In orthodox quantum mechanics, a system's wave function governs the probabilities for the outcomes of experiments on the system, whereas in GRWm/GRWf, the wave function governs the PO. This difference creates a difference concerning what one means when saying that the cat is dead: In orthodox quantum mechanics, one means that if we made a ``quantum measurement'' of the cat's macro-state, we would with probability 1 find it dead, whereas in GRWm/GRWf one means that the PO forms a dead cat. If the cat is dead in this sense, in GRWm/GRWf, and $\psi$ is close but not exactly equal to $\dead$, then there is still a tiny but non-zero probability that within the next millisecond the collapses occur in such a way that the cat is suddenly alive! But that does not contradict the claim that a millisecond before the cat was dead; it only means that GRWm/GRWf allows resurrections to occur---with tiny probability! In particular, if we observe the cat after that millisecond, there is a positive probability that we find it alive (simply because it \emph{is} alive) even though before the millisecond it actually was dead.\footnote{For GRWf, I assume here that the reason why $c_2$ is small lies in GRW collapses that have occurred in the past. When that is not the case, for example when $\psi$ is prepared to be \eqref{cat} with small $c_2$, the situation in GRWf is yet different, as then there is no fact at the initial time from the PO as to whether the cat is dead or is alive. This will play a role in the subsequent paradox.}

\subsection{Paradox 3: Consider Many Systems}

Paradox: A variant of the previous paradox was formulated by \citet{Lew99} in terms of counting marbles; the discussion continued in \citep{Cor99,CM99,CM00,BG99a,BG99b,BG01,Fri02, Lew03,Lew05,Lew06, Mon04,Wal08,Pea06,McQu15}. Let $\psi_1$ be the state ``the marble is inside the box'' and $\psi_2$ the state ``the marble is outside the box''; these wave functions have disjoint supports $S_1,S_2$ in configuration space (i.e., wherever one is nonzero the other is zero). Let $\psi$ be given by \eqref{cat} with $0<|c_2|^2 \ll |c_1|^2<1$; finally, consider a system of $n$ (non-interacting) marbles at time $t_0$, each with wave function $\psi$, so that the wave function of the system is $\psi^{\otimes n}$. Then, for each of the marbles we would feel entitled to say that it is inside the box, but on the other hand, the probability that all marbles be found inside the box is $|c_1|^{2n}$, which can be made arbitrarily small by making $n$ sufficiently large.

\medskip

Answer: For GRWm it follows from the PO, as in the answer to the previous paradox, that each of the marbles is inside the box at the initial time $t_0$. However, it is known that a superposition like \eqref{cat} of macroscopically distinct states $\psi_i$ will approach under the GRW evolution either a wave function $\psi_1(\infty)$ concentrated in $S_1$ or another $\psi_2(\infty)$ in $S_2$ with probabilities $|c_1|^2$ and $|c_2|^2$, respectively. (Here I am assuming $H=0$ for simplicity. Although both coefficients will still be nonzero after any finite number of collapses, one of them will tend to zero in the limit $t\to\infty$.) Thus, for large $n$ the wave function will approach one consisting of approximately $n|c_1|^2$ factors $\psi_1(\infty)$ and $n|c_2|^2$ factors $\psi_2(\infty)$, so that ultimately about $n|c_1|^2$ of the marbles will be inside and about $n|c_2|^2$ outside the box---independently of whether anybody observes them or not. The occurrence of some factors $\psi_2(\infty)$ at a later time provides another example of the resurrection-type events mentioned above; they are unlikely but do occur, of course, if we make $n$ large enough. 

The act of observation plays no role in the argument and can be taken to merely record pre-existing macroscopic facts. To be sure, the physical interaction involved in the act of observation may have an effect on the system, such as speeding up the evolution from $\psi$ towards either $\psi_1(\infty)$ or $\psi_2(\infty)$; but GRWm provides unambiguous facts about the marbles also in the absence of observers.

In GRWf, the story is a little more involved. The fact that the answer depends on the choice of PO illustrates again the relevance of the PO, as well as the necessity to make the PO and its laws explicit. The story is more involved because the PO cannot be considered at only one point in time $t_0$ but needs to be considered over some time interval (say a millisecond), and because it depends on randomness. First, if we assume that the smallness of $c_2$ is due to previous collapses centered inside the box, then the flashes during the millisecond before $t_0$ form $n$ marbles inside the box. Thus, as in GRWm, initially we have $n$ marbles inside the box, of which $n|c_2|^2$ will be outside the box after a while.

Now consider the other case: that the smallness of $c_2$ is not due to previous collapses, but due to some other method of preparing $\psi$. Then we may have to consider only flashes after $t_0$. Consider first a single marble. Something improbable may already happen at this stage; for example, all the flashes might occur outside the box. In that case we would say that the marble \emph{is} outside the box. As well, it might happen that half of the flashes occur outside and half of them inside the box; in that case we would say that half of the marble's matter is located inside the box. The overwhelmingly probable case, of course, is that more than 99\%\ of the marble's flashes occur inside the box, a case in which it is reasonable to say that the marble is inside the box. Thus, if $|c_2|^2 \ll |c_1|^2$ at time $t_0$ then in GRWf (unlike in GRWm) the marble is not necessarily (only very probably) inside the box, provided the time interval we consider is the millisecond after $t_0$. Now consider $n$ marbles, with $n$ so big that $|c_1|^{2n}\ll 1$; then it is not probable any more that for \emph{all} marbles 99\%\ of the flashes occur inside the box. Rather, the overwhelmingly probable case is that for the majority of marbles 99\%\ of the flashes occur inside the box (so that one should say these marbles are inside), while for a few marbles a significant fraction of flashes occurs outside, and for extremely few marbles even all flashes occur outside. In the limit $t\to\infty$, as a consequence of the convergence to either $\psi_1(\infty)$ or $\psi_2(\infty)$, for each marble either almost all flashes occur inside the box or almost all flashes occur outside.

\subsection{What is Real and What is Accessible}

Another remark concerns the reply by \citet{BG99a,BG99b,BG01} (also \citeyear{BG03}, Sec.~11) to the tail problem and the marble problem; see also \citep{Mon04}. Bassi and Ghirardi use GRWm but immediately focus on (what they call) the \emph{accessibility} of the matter density, i.e., on the fact that the matter density cannot be measured by inhabitants of the GRWm world to arbitrary accuracy. In particular, they point out that for the marble wave function the matter density outside the box is not detectable. I think that Bassi and Ghirardi took two steps at once, thereby making the argument harder to understand for their readers, and that perhaps they did not take the PO seriously enough. To the extent that the worry expressed in the above paradoxes (and by \citet{Shi,AL90,Lew99,CM99}) is whether GRW theories do give rise to unambiguous facts about the aliveness of Schr\"odinger's cat or the location of the marble, it concerns whether GRW theories provide a picture of reality that conforms with our everyday intuition. Such a worry cannot be answered by pointing out what an observer can or cannot measure. Instead, I think, the answer can only lie in what the ontology \emph{is like}, not in what observers see of it. Moreover, I think, it can only lie in what the \emph{primitive ontology} is like, as that is the part of the ontology representing ``the cat'' and ``the marble.'' It can easily cause confusion to blur the  distinction between what is real and what is accessible, as not everything that is real and well-defined according to GRW theories is accessible; for example, the number of collapses in a given time interval $[t_1,t_2]$ is well-defined but cannot be measured reliably \citep{CT14,CT13}.

Now for the marble state considered above, it is a fact for the matter density ontology that the fraction $|c_2|^2 \ll 1$ of the marble's matter lies outside the box---a fact that does not contradict the claim that the marble is inside the box, as can be illustrated by noting that anyway, for thermodynamic reasons, the marble creates a vapor out of some of its atoms (with low partial pressure), an effect typically outweighing $|c_2|^2$. The state of the PO in which the overwhelming majority of matter is inside the box justifies saying that the marble is inside the box. Thus, the PO does provide a picture of reality that conforms with our everyday intuition. All this is independent of whether the PO is observable (accessible) or not. Bassi and Ghirardi sometimes sound as if they did not take the matter density seriously when it is not accessible; I submit that the PO should always be taken seriously. 

The reason why Bassi and Ghirardi attribute such importance to whether the matter density is accessible is presumably the following: If the matter density outside the box could be measured to be nonzero then this would seem to threaten the claim that the marble is inside. But the threat is actually not serious: For example, the vapor created by the marble can in fact be measured to have nonzero density, but this fact does not at all suggest that ``the marble'' should be regarded as being outside, it only suggests (indeed, it entails) that a small part of the marble's matter is outside the box.

\section{The Need for a Primitive Ontology}
\label{sec:need}

In this section, I offer further considerations about the status and role of the PO.

\subsection{Is There a Cat?}

We all once learned OQM and got used to ``quantum think''---certain ways of reasoning employed in OQM. One of these ways of reasoning postulates that, whenever $\psi$ is the wave function of a live cat,\footnote{I.e., a wave function concentrated in a certain region of configuration space, and with energies of certain subsystems in suitable ranges, etc..} there is a live cat. Since collapse theories are not OQM, these ways of reasoning need not apply, and this postulate is questionable. Bohmian mechanics provides an example of a situation in which the PO could be in a configuration of a live cat, and move with time in the manner of a live cat, while $\psi$ is a non-trivial superposition of a live cat and a dead cat. This example suggests that 
\be\label{S3}
\text{\begin{minipage}{11cm}
it is appropriate to say ``there is a live cat'' when and only when part of the PO behaves like a live cat.
\end{minipage}} 
\ee
In particular, whether there is a live cat depends on $\psi$ only through the dependence of the PO on $\psi$ by virtue of the laws governing the PO. It then follows that there is no live cat in GRW$\emptyset$, because there is no PO, regardless of $\psi$. This is perhaps the most fundamental problem about theories without a PO: It is hard to avoid the conclusion that they imply the absence of matter in space-time.

Furthermore, \eqref{S3} implies that there is a logical gap between saying
\be\label{S1}
 \text{``$\psi$ is the wave function of a live cat''} 
\ee
and saying 
\be\label{S2}
\text{``there is a live cat.''}
\ee
After all, in Bohmian mechanics, \eqref{S2} follows from \eqref{S1} by virtue of a law of the theory, which asserts that the configuration $Q(t)$ of the PO is $|\psi_t|^2$ distributed at every time $t$. This law entails that if $\psi_t$ is the wave function of a live cat, then $Q(t)$ will be the configuration of a live cat and will continue to behave like the configuration of a live cat for $t'>t$ as long as $\psi_{t'}$ stays the wave function of a live cat. Thus, Bohmian mechanics suggests that \eqref{S2} would not follow from \eqref{S1} if there was not a law connecting the two by means of the PO.

An attractive feature of the PO view is that the word ``cat'' actually refers to a \emph{thing}: a set of particles in Bohmian mechanics, or a set of flashes in GRWf, or a part of the continuous matter in GRWm that behaves like a cat under the influence of $\psi$. Correspondingly, the statement ``the cat is alive'' simply means that that \emph{thing} behaves in a certain way---like a live cat. In GRW$\emptyset$, in contrast, we need to assume a re-interpretation of English phrases: We need to assume that the word ``cat'' does not refer to any \emph{thing}, and that the phrase ``the cat is alive'' does not summarize the behavior of a thing called ``the cat,'' but rather that this phrase really means \eqref{S1}. I find that hard to swallow.

Despite all this, there is a sense in which GRW$\emptyset$ works: The GRW wave function $\psi_t$ is, at almost all times, concentrated, except for tiny tails, on a set of configurations that are macroscopically equivalent to each other. Thus, we can read off from $\psi_t$ what the macro-state is: For example, we could pretend there were particles whose configuration $Q$ is $|\psi_t|^2$ distributed, and consider the macroscopic appearance of $Q$. Alternatively, we could compute $m(x,y,z,t)$ and consider its macroscopic appearance. Some thought reveals that the macroscopic appearance of $Q$ agrees (with high probability) with that of $m$ [and in fact with that of the flashes in GRWf \citep{GTZ12}; but see also \citep{Seb15}]; for example, they agree about which way the pointer of a measurement instrument points. So $\psi_t$ contains all the information about what the macro-configuration would look like, if there were any matter that could have this macro-configuration. This fact explains why GRW$\emptyset$ provides enough information to read off empirical predicitions, although they are, in fact, the empirical predictions of GRWf and GRWm, and not those of GRW$\emptyset$.

Actually, while GRWf and GRWm are empirically equivalent in the sense that there is no experiment that could test one against the other, their macro-configurations of the PO can differ if the parameters (i.e., constants of nature) of the GRW time evolution (i.e., the collapse width $\sigma$ and the collapse rate $\lambda$) are chosen in extreme ways. This became visible in \citep{FT12}, where we drew a parameter diagram with axes $\sigma$ and $\lambda$ for the GRW theories and therein the empirically refuted region and the philosophically unsatisfactory region: it turned out that the latter region is different for GRWf than for GRWm if based on the following definition:
``We regard a parameter choice $(\sigma,\lambda)$ as philosophically satisfactory if and only if the PO agrees on the macroscopic scale with what humans normally think macroscopic reality is like.''~\citep{FT12} (But see also \citep{Seb15} for a further subtlety about the two regions.)

I remark that an issue parallel to GRW$\emptyset$ versus GRWm/GRWf comes up in many-worlds theories. The most popular understanding of many-worlds is that there is only the wave function $\psi$ which evolves according to the linear Schr\"odinger equation. The theory actually becomes clearer if one introduces a suitable PO; an example is provided by Schr\"odinger's many-worlds theory \citep{mw}.

\subsection{All Observables?}
\label{sec:all}

\citet{Pea06} suggested, instead of GRWm, the view that \emph{every} observable $\hat{A}$ is attributed as a ``true value'' the value that in OQM would be the average of $\hat{A}$ in the state $\psi_t$, 
\be\label{Atdef}
A(t) = \scp{\psi_t}{\hat A|\psi_t}\,,
\ee
and that the $m(x,y,z,t)$ function of GRWm is just the special case for $\hat{A}=\hat{M}(x,y,z)$ the mass density operator, which in the position representation of non-relativistic QM of $N$ particles is multiplication by the function
\be
\sum_{i=1}^N m_i\, \delta(x_i-x) \, \delta(y_i-y)\, \delta(z_i-z)\,.
\ee
To contrast this view with GRWm, let me use the creation myth again that I mentioned in Section~\ref{sec:PO} above: For creating a GRWm world, God would need two acts of creation, one for creating $\psi$ and one for creating the matter with density $m(x,y,z,t)$. For creating a world according to Pearle's view, would God just have to create $\psi$, or would he have to carry out a separate act of creation for every observable $\hat{A}$? Let us compare the situation to the status of energy in a Newtonian universe. If God wants to create a Newtonian universe, he needs to create point particles that move according to the Newtonian equation of motion,
\be
m_i \, \frac{d^2\vQ_i(t)}{dt^2} = - \nabla_i V\Bigl(\vQ_1(t),\ldots,\vQ_N(t) \Bigr)
\ee
for a suitable potential $V$. Then energy is defined as
\be
E(t) = \sum_i \frac{m_i}{2} \Bigl|\frac{d\vQ_i}{dt}\Bigr|^2 + V\Bigl(\vQ_1(t),\ldots,\vQ_N(t) \Bigr)\,,
\ee
and no separate act of creation is needed in which God would have to create energy according to this equation. This situation suggests that likewise, no separate act of creation is needed to introduce $A(t)$. So Pearle's suggestion is really a form of GRW$\emptyset$, notwithstanding the appearance that GRWm be contained in it.

Let me connect Pearle's suggestion to empricial predictions. It is one of the key roles of the PO that when deriving empirical predictions (``with probability $p$, the pointer will be in position $x$''), we actually derive statements about the PO, as the pointer is where its flashes/matter density/particles are. It is illuminating to see how this works out in various theories with PO \citep{AGTZ14}. In Pearle's approach, in contrast, if we use the $m$ function to determine the pointer position, then this is just one among many possibilities of defining what is meant by ``pointer position.'' Other possibilities could be based on the flash ontology, or on taking $\hat{A}$ to be the center-of-mass position operator of the particles forming the pointer. It is not clear, then, which choice would be correct in case they disagree [and, as mentioned above, GRWf and GRWm sometimes do disagree about the macro-configuration of the PO for extreme values of $\sigma$ and $\lambda$ \citep{FT12}]. But even when they agree, I find it unclear what justifies these choices and prefers them over all other operators $\hat{A}$. It seems within our power to define which pieces of matter we call pointers, but not to define what is meant by their position.

\subsection{Paradoxes}

The paradoxes of Section~\ref{sec:paradoxes} also pose a problem for GRW$\emptyset$. For example, consider Paradox 1 again, the most basic one: After Schr\"odinger's cat collapses to $|$alive$\rangle$, the amplitude of $|$dead$\rangle$ is tiny but not exactly zero. What right then do we have to say that the cat \emph{is alive}? To solve this problem, \citet{AL96} have proposed a rule under the name ``fuzzy link'' asserting that any observable $\hat{A}$ has value $A$ whenever $\psi$ lies approximately in the eigenspace of $\hat{A}$ with eigenvalue $A$. Applying this rule to $\hat{A}$ the projection to the space of wave functions of live cats yields that whenever the amplitude of $|$dead$\rangle$ in a superposition of $|$alive$\rangle$ and $|$dead$\rangle$ is tiny, the cat is alive. The question arises, however, why we would have the freedom to introduce a rule such as the fuzzy link. When we formulate a fundamental physical theory, we have the freedom to posit which things exist according to the theory (the ontology) and the laws governing them. For GRW$\emptyset$, we would posit that $\psi$ and only $\psi$ exists, and that it evolves according to the GRW process. It seems that there is no further room, then, for additional postulates such as the fuzzy link. Of course, if we introduce further elements of the ontology, such as a PO, then we can also postulate further laws governing the further ontology.

\subsection{Worlds in the Tails}
\label{sec:mw}

Another puzzle about the tails \citep{Cor99,Wal08,Mau10,McQu15,EE15} arises from the fact that in GRWm, there is some matter that is governed by the tails of $\psi$: The $m$ function has small ripples that follow the evolution of suppressed parts of $\psi$, for example of a dead cat after $2^{-1/2}(\dead + \alive)$ has collapsed to near $\alive$. Arguably, as in Schr\"odinger's many-worlds theory with a PO \citep{mw}, this means that GRWm has a many-worlds character: Apart from the live cat, there also exists a dead cat. In GRWf, in contrast, when the tails are too small, there are no flashes associated with them, so that such further worlds do not exist. In GRWm, one of the worlds is the ``fat world'' (associated with the bulk of the matter), the others are the ``faint worlds'' (associated with the tails). I think we are forced to admit that the faint worlds are real. As pointed out by \citet{Wal14}, the faint worlds behave very differently than the fat world, in fact catastrophically, due to the gradient in the Gaussian that gets multiplied onto $\psi$ at every collapse. It seems that the upshot is that GRWm predicts the existence of several worlds, the fat and the faint ones, with very different behavior, and that we live in the fat world (as most observers do, since observers do not survive for long in the faint worlds).

\subsection{Lorentz Invariance}

Another consideration concerns Lorentz invariance. Relativistic versions of GRWf and GRWm have been developed \citep{Tum06,Tum06c,BDGGTZ} for non-interacting particles; relativistic collapse processes for $\psi$ (albeit ultraviolet divergent) had been developed even earlier \citep{Pea90,Dio90}. The flashes transform like space-time points and the $m$ function like a Lorentz scalar, while there is no simple relation between $\psi_\Sigma$ and $\psi_{\Lambda(\Sigma)}$ for spacelike hypersurfaces $\Sigma$ and Lorentz transformations $\Lambda$. For GRW$\emptyset$, in contrast, it seems unsatisfactory to merely replace the GRW process by a Lorentz invariant collapse process, for the following reason \citep{BDGGTZ}. For a spacelike 3-surface $\Sigma=A\cup B$ with $A\cap B=\emptyset$, the reduced density matrix of $A$ is
\be
\hat\rho_A = \tr_B \pr{\psi_\Sigma}\,.
\ee
But this depends on $B$: For $\Sigma'=A\cup B'$, $\psi_{\Sigma'}$ may be very different from $\psi_{\Sigma}$ due collapses between $\Sigma$ and $\Sigma'$, and $\hat\rho_A'=\tr_{B'} \pr{\psi_{\Sigma'}}$ may be very different from $\hat\rho_A$. For example, if $\psi_\Sigma=2^{-1/2}(|\!\!\uparrow\downarrow\rangle - |\!\!\downarrow\uparrow\rangle)$ is the spin-singlet state of an EPR pair and Bob's particle passes through a Stern--Gerlach magnet and collapses to either $|\!\!\uparrow\downarrow\rangle$ or $|\!\!\downarrow\uparrow\rangle$ before reaching $\Sigma'$, then the reduced state on Alice's side is $\hat\rho_A=\tfrac{1}{2}(|\!\!\uparrow\rangle\langle\uparrow \!\!|+|\!\!\downarrow\rangle\langle \downarrow\!\!|)$ before the collapse and either $\hat\rho_A'=|\!\!\uparrow\rangle\langle\uparrow\!\!|$ or $\hat\rho_A'=|\!\!\downarrow\rangle\langle \downarrow\!\!|$ afterward. Now, in GRW$\emptyset$ we would have liked to read off the local state of affairs in region $A$ from $\psi$, but the answer depends on the choice of $B$ or $B'$. However, local facts in region $A$ should not depend whether we consider $B$ or $B'$! This problem is absent in GRWf and GRWm because, even though reduced density matrices are still subject to the same ambiguity, the PO consists of local variables (the flashes or matter density) and thus provides plenty of local facts, enough for a meaningful physical theory. Albert (personal communication, 2014) has suggested that for GRW$\emptyset$, one could simply use a preferred foliation of space-time into spacelike hypersurfaces. This would have the disadvantage of making GRW$\emptyset$ less relativistic than GRWf or GRWm, but the bigger problem with this suggestion is that, as with the fuzzy link rule, there does not seem to be any room for further postulates (such as to use a particular foliation for extracting local facts) after it has been stipulated that $\psi$ evolves according to a certain relativistic collapse law and that the ontology contains only $\psi$.

\section{Connection to the Mind-Body Problem}
\label{sec:mind}

The persistent disagreement between two camps, one maintaining that GRW$\emptyset$ is an acceptable theory [e.g., \citet{Adl07,Alb15,Pea06}; Rimini], and the other that a PO is needed [e.g., \citet{All13a,All13b,All15a,All15b,BGG95,BDGGTZ,GTZ12,Mau10,Mau14}], is perhaps ultimately related to different views about the mind-body problem. After a brief outline of this problem [for a detailed description see, e.g., the first few chapters of \citep{Ch96}], I will explain how it comes up in connection with requirements on fundamental physical theories (such as, to have a PO).

\subsection{What is the Mind-Body Problem?} 

The color red looks a particular way to me that I cannot express in words. I would guess it looks the same way to other people, but I cannot really check (after all, it is logically possible that things that look red to me, such as tomatoes, look blue to somebody else and that we would never notice because that other person will call things red when they look blue to her). ``The way red looks'' is the conscious experience of red, which is different from the knowledge that the light that reached my eye had a wave length of 800 nanometers and is called ``red'' in English. 

If we try to come up with an explanation of this experience, we encounter a problem: A physical theory may claim that the particles in my brain follow certain trajectories, but this claim would not explain the experience of red, regardless of what the trajectories are. This is the essence of the mind-body problem, also sometimes called the ``hard problem of consciousness.'' In contrast, all the information processing (from light of 800 nm to the English word ``red'') poses no obstacle in principle for an explanation in terms of particle trajectories. (It is the ``easy problem of consciousness.'') The mind-body problem is not specific to particles, it would be the same with fields, wave functions, or any other kind of physical object. Here, consciousness does not mean being awake, or being aware, or knowing oneself; but seeing colors.

To appreciate the problem, it may help to try to write a computer program that makes the computer see red. That seems impossible. It is clearly possible that the computer identifies data from a digital camera as corresponding to a wave length of 800 nanometers and to the English word ``red.'' However, for this information processing, the computer need not see red, and by virtue of this information processing alone, the computer does not see red.

However, there is no consensus about the mind-body problem. Functionalists think that there is no such problem, and that there is nothing more to the mind than information processing. I think there is a mind-body problem, and I do not see how it could ever be solved, i.e., how conscious experiences could be reduced to physical processes. I conclude that there are facts in the world beyond the physical facts.

\subsection{How the Mind-Body Problem Affects Physics}

A goal of a fundamental physical theory is to explain our experiences. In order to even connect with our experiences, it may seem that the theory has to solve the mind-body problem, which seems hopeless. 

Luckily, there is a simple way out (as emphasized by Maudlin). Suppose the theory implies the existence of macroscopic objects in 3-space in particular macroscopic configurations. And suppose that we are not completely deluded about the world around us, that instrument pointers actually point more or less the way we think, and that we can usually read correctly. Then we can compare the predicted macro-configurations to the perceived macro-configurations, and claim, in case of agreement, that the theory is empirically adequate. It actually works like that in classical physics, and in any theory with a primitive ontology.

Here is a passage from \citet{Mau14} in this direction:
\begin{quotation}
A theory's specification of the fermion density [as an example of a PO] in every region of the universe entails the distribution of matter at macroscopic scale. And if what it predicts at macroscopic scale matches everything we think we know about the world at macroscopic scale (including where the pointers ended up pointing, where the ink is on the paper, the shape of the earth, the dimensions of the Empire State building, etc., etc., etc.)\ then the theory is empirically adequate in any reasonable sense. There may be objections to such a theory, but they cannot rightfully be called empirical objections.
\end{quotation}

That is, physical theories with a PO do not need to solve the mind-body problem, whereas theories without a PO seem to be stuck with this unsolvable problem. Thus, physical theories are better off with a PO.

\bigskip

\noindent\textit{Acknowledgments.} I thank David Albert, Eddy Chen, Shelly Goldstein, Ned Hall, Travis Norsen, Philip Pearle, Antoine Tilloy, and Nino Zangh\`\i\ for helpful discussions.


\begin{thebibliography}{29}

\bibitem[Adler(2007)]{Adl07} S. L. Adler:
	Lower and Upper Bounds on CSL Parameters from Latent Image
	Formation and IGM Heating.
	\textit{Journal of Physics A: Mathematical and Theoretical} \textbf{40}: 2935--2957 (2007)
	\url{http://arxiv.org/abs/quant-ph/0605072}

\bibitem[Albert(2015)]{Alb15} D. Albert: 
	\textit{After Physics.}
	Harvard University Press (2015)

\bibitem[Albert and Loewer(1990)]{AL90} D. Albert and B. Loewer:
	Wanted Dead or Alive: Two Attempts to Solve Schr\"odinger's Paradox.
	Pages 277--285 in A. Fine, M. Forbes and L. Wessels
	(ed.s), \textit{Proceedings of the 1990 Biennial Meeting of the
	Philosophy of Science Association. Vol. 1}. 
	East Leasing: Philosophy of Science Association (1990)

\bibitem[Albert and Loewer(1996)]{AL96} D. Albert and B. Loewer:
	Tails of Schr\"odinger's Cat.
	Pages 81--91 in R. Clifton (ed.), \textit{Perspectives on Quantum Reality}.
	Dordrecht: Kluwer (1996)
	
\bibitem[Allori(2013a)]{All13a} V. Allori: 
	On the Metaphysics of Quantum Mechanics.
	Pages 116--140 in S. LeBihan (ed.):
	\textit{Pr\'ecis de philosophie de la physique}. 
	Paris: Vuibert (2013a)
	\url{http://philsci-archive.pitt.edu/9343/}

\bibitem[Allori(2013b)]{All13b} V. Allori: 
	Primitive Ontology and the Structure of Fundamental Physical Theories.
	Pages 58--75 in D. Albert and A. Ney (ed.s), 
	\textit{The Wave Function.} 
	Oxford University Press (2013b)
	\url{http://philsci-archive.pitt.edu/9342/}
	
\bibitem[Allori(2015a)]{All15a} V. Allori:
	Primitive Ontology in a Nutshell.
	{\it International Journal of Quantum Foundations} {\bf 1}: 107--122 (2015a)
	\url{http://philsci-archive.pitt.edu/11651/}

\bibitem[Allori(2015b)]{All15b} V. Allori:
	How to Make Sense of Quantum Mechanics (and More): Fundamental Physical 
	Theories and Primitive Ontology.
	Preprint (2015b)
	\url{http://philsci-archive.pitt.edu/11652/}

\bibitem[Allori et al.(2008)Allori, Goldstein, Tumulka, and Zangh\`\i]{AGTZ06} V. Allori, S. Goldstein, R. Tumulka, and N. Zangh\`\i:
	On the Common Structure of Bohmian Mechanics and the Ghirardi--Rimini--Weber Theory. 
	\textit{British Journal for the Philosophy of Science} \textbf{59}:  353--389 (2008)
	\url{http://arxiv.org/abs/quant-ph/0603027}

\bibitem[Allori et al.(2011)Allori, Goldstein, Tumulka, and Zangh\`\i]{mw} V. Allori, S. Goldstein, R. Tumulka, and N. Zangh\`\i:
	Many-Worlds and Schr\"odinger's First Quantum Theory.
	\textit{British Journal for the Philosophy of Science} \textbf{62(1)}: 1--27 (2011)
	\url{http://arxiv.org/abs/0903.2211}

\bibitem[Allori et al.(2014)Allori, Goldstein, Tumulka, and Zangh\`\i]{AGTZ14} V. Allori, S. Goldstein, R. Tumulka, and N. Zangh\`\i: 
	Predictions and Primitive Ontology in Quantum Foundations: A Study of Examples.
	\textit{British Journal for the Philosophy of Science} \textbf{65}: 323--352 (2014)
	\url{http://arxiv.org/abs/1206.0019}

\bibitem[Bassi and Ghirardi(1999a)]{BG99a} A. Bassi and G.C. Ghirardi: 
	Do dynamical reduction models imply that arithmetic does not apply to 
	ordinary macroscopic objects?
	\textit{British Journal for the Philosophy of Science} \textbf{50}: 49--64 (1999a)
	\url{http://arxiv.org/abs/quant-ph/9810041}

\bibitem[Bassi and Ghirardi(1999b)]{BG99b} A. Bassi and G.C. Ghirardi:
	More about Dynamical Reduction and the Enumeration Principle.
    	\textit{British Journal for the Philosophy of Science} \textbf{50}: 719 (1999b)
	\url{http://arxiv.org/abs/quant-ph/9907050}

\bibitem[Bassi and Ghirardi(2001)]{BG01} A. Bassi and G.C. Ghirardi:
	Counting Marbles: Reply to Clifton and Monton.
	\textit{British Journal for the Philosophy of Science} \textbf{52}: 125--130 (2001)

\bibitem[Bassi and Ghirardi(2003)]{BG03} A. Bassi and G.C. Ghirardi: 
	Dynamical Reduction Models.
	\textit{Physics Reports} \textbf{379}: 257--426 (2003)
	\url{http://arxiv.org/abs/quant-ph/0302164}

\bibitem[Bassi and Ulbricht(2014)]{BU14} A. Bassi and H. Ulbricht:
	Collapse models: from theoretical foundations to experimental verifications.
	\textit{Journal of Physics: Conference Series} \textbf{504}: 012023 (2014)
	\url{http://arxiv.org/abs/1401.6314}

\bibitem[Bedingham et al.(2014)Bedingham, D\"urr, Ghirardi, Goldstein, Tumulka, and Zangh\`\i]{BDGGTZ} D. Bedingham, D. D\"urr, G.C. Ghirardi, S. Goldstein, R. Tumulka, and N. Zangh\`\i:
	Matter Density and Relativistic Models of Wave Function Collapse.
	\textit{Journal of Statistical Physics} \textbf{154}: 623--631 (2014)
	\url{http://arxiv.org/abs/1111.1425}

\bibitem[Bell(1976)]{Bell76} J. S. Bell:
	The theory of local beables.
	\textit{Epistemological Letters} \textbf{9}: 11 (1976).	
	Reprinted as chapter 7 of \citep{Bell87b}.

\bibitem[Bell(1980)]{Bell80} J.~S.~Bell: 
	De Broglie--Bohm, Delayed-Choice Double-Slit Experiment, and Density Matrix.
	\textit{International Journal of Quantum Chemistry} \textbf{14}: 155--159 (1980). 
	Reprinted as chapter 14 of \citep{Bell87b}.

\bibitem[Bell(1987a)]{Bell87} J. S. Bell: 
	Are There Quantum Jumps? 
	Pages 41--52 in C. W. Kilmister (ed.),
	\textit{Schr\"odinger. Centenary Celebration of a Polymath}.
	Cambridge University Press (1987a). 
	Reprinted as chapter 22 of \citep{Bell87b}.
  
\bibitem[Bell(1987b)]{Bell87b} J. S. Bell: 
	\textit{Speakable and Unspeakable in Quantum Mechanics}. 
	Cambridge University Press (1987b)

\bibitem[Ghirardi et al.(1995)Benatti, Ghirardi, and Grassi]{BGG95} F. Benatti, G.C. Ghirardi, and R. Grassi: 
	Describing the macroscopic world: closing the circle within
	the dynamical reduction program.
	\textit{Foundations of Physics} {\bf 25}: 5--38 (1995)

\bibitem[Bernstein(2011)]{Ber13} J. Bernstein: 
	More About Bohm's Quantum.
	\textit{American Journal of Physics} \textbf{79}: 601 (2011)

\bibitem[Carlesso et al.(2016)Carlesso, Bassi, Falferi, and Vivante]{CBFV16} M. Carlesso, A. Bassi, P. Falferi, and A. Vinante:
	Experimental bounds on collapse models from gravitational wave detectors.
	{\it Physical Review D} {\bf 94}: 124036 (2016)
	\url{http://arxiv.org/abs/1606.04581}

\bibitem[Chalmers(1996)]{Ch96} D. Chalmers:
	\emph{The Conscious Mind.}
	Oxford University Press (1996)

\bibitem[Clifton and Monton(1999)]{CM99} R. Clifton and B. Monton:
	Losing your marbles in wavefunction collapse theories.
	\textit{British Journal for the Philosophy of Science} \textbf{50}: 697--717 (1999)
	\url{http://arxiv.org/abs/quant-ph/9905065}

\bibitem[Clifton and Monton(2000)]{CM00} R. Clifton and B. Monton:
	Counting marbles with `accessible' mass density: a reply to Bassi and Ghirardi.
	\textit{British Journal for the Philosophy of Science} \textbf{51}: 155--164 (2000)
	\url{http://arxiv.org/abs/quant-ph/9909071}
    
\bibitem[Colin et al.(2006)Colin, Durt, and Tumulka]{CDT05} S. Colin, T. Durt, and R. Tumulka:
	On Superselection Rules in Bohm--Bell Theories.
	\textit{Journal of Physics A: Mathematical and General} \textbf{39}: 15403--15419 (2006)
	\url{http://arxiv.org/abs/quant-ph/0509177}
  
\bibitem[Cordero(1999)]{Cor99} A. Cordero: 
	Are GRW Tails as Bad as They Say?
	\textit{Philosophy of Science} \textbf{66}: S59--S71 (1999)

\bibitem[Cowan and Tumulka(2014)]{CT14} C.~W.~Cowan and R.~Tumulka:
	Can One Detect Whether a Wave Function Has Collapsed?
	\textit{Journal of Physics A: Mathematical and Theoretical} \textbf{47}: 195303 (2014)
	\url{http://arxiv.org/abs/1307.0810}

\bibitem[Cowan and Tumulka(2016)]{CT13} C. W. Cowan and R. Tumulka:
	Epistemology of Wave Function Collapse in Quantum Physics.
	\textit{British Journal for the Philosophy of Science} {\bf 67(2)}: 405--434 (2016)
	\url{http://arxiv.org/abs/1307.0827}

\bibitem[Di\'osi(1989)]{Dio89} L. Di\'osi:
	Models for universal reduction of macroscopic quantum fluctuations. 
	\textit{Physical Review A} \textbf{40}: 1165--1174 (1989)

\bibitem[Di\'osi(1990)]{Dio90} L. Di\'osi:
	Relativistic theory for continuous measurement of quantum fields.
	\textit{Physical Review A} \textbf{42}: 5086--5092 (1990)

\bibitem[D\"urr et al.(2004)D\"urr, Goldstein, and Zangh\`\i]{DGZ04} D.~D\"urr, S.~Goldstein, and N.~Zangh\`\i: 
	Quantum Equilibrium and the Role of Operators as Observables in Quantum Theory.
	{\it Journal of Statistical Physics} {\bf 116}: 959--1055 (2004)
	\url{http://arxiv.org/abs/quant-ph/0308038}

\bibitem[Egg and Esfeld(2015)]{EE15} M. Egg and M. Esfeld:
	Primitive ontology and quantum state in the GRW matter density theory.
	{\it Synthese} {\bf 192}:  3229--3245 (2015)
	\url{http://philsci-archive.pitt.edu/11102/}

\bibitem[Esfeld(2014)]{Esf14} M. Esfeld:
	The primitive ontology of quantum physics: guidelines for an assessment of the proposals.
	\textit{Studies in History and Philosophy of Modern Physics} \textbf{47}: 99--106 (2014)
	\url{http://philsci-archive.pitt.edu/10711/}

\bibitem[Feldmann and Tumulka(2012)]{FT12} W. Feldmann and R. Tumulka: 
	Parameter Diagrams of the GRW and CSL Theories of Wave Function Collapse.
	\textit{Journal of Physics A: Mathematical and Theoretical} \textbf{45}: 065304 (2012)
	\url{http://arxiv.org/abs/1109.6579}

\bibitem[Frigg(2003)]{Fri02} R. Frigg: 
	On the Property Structure of Realist Collapse Interpretations of Quantum Mechanics 
	and the So-Called ``Counting Anomaly.''
	\textit{International Studies in the Philosophy of Science} \textbf{17}: 43--57 (2003)
	\url{http://sas-space.sas.ac.uk/1041/}
	
\bibitem[Ghirardi(2007)]{Ghi07} G.C. Ghirardi: Collapse Theories. 
	In E. N. Zalta (ed.), \textit{Stanford Encyclopedia of Philosophy}, 
	published online by Stanford University. 
	\url{http://plato.stanford.edu/entries/qm-collapse/}
	(2007)

\bibitem[Ghirardi et al.(1986)Ghirardi, Rimini, and Weber]{GRW86} G.C. Ghirardi, A. Rimini, and T. Weber: 
	Unified Dynamics for Microscopic and Macroscopic Systems. 
	\textit{Physical Review D} \textbf{34}: 470--491 (1986)

\bibitem[Gisin(1984)]{Gis84} N. Gisin:
	Quantum Measurements and Stochastic Processes.
	 {\it Physical Review Letters} {\bf 52}: 1657 (1984)

\bibitem[Gisin and Percival(1993)]{GP93} N. Gisin and I. C. Percival:
        The quantum state diffusion picture of physical processes.
        \textit{Journal of Physics A: Mathematical and General} \textbf{26}: 2245--2260 (1993)

\bibitem[Goldstein(1998)]{Gol98} S. Goldstein: 
	Quantum Theory Without Observers.
	\textit{Physics Today}, Part One: March 1998, 42--46. Part Two: April 1998, 38--42.

\bibitem[Goldstein et al.(2012)Goldstein, Tumulka, and Zangh\`\i]{GTZ12} S. Goldstein, R. Tumulka, and N. Zangh\`\i:
	The Quantum Formalism and the GRW Formalism.
	\textit{Journal of Statistical Physics} \textbf{149}: 142--201 (2012)
	\url{http://arxiv.org/abs/0710.0885}

\bibitem[Leggett(2002)]{Leg02} A. J. Leggett: 
	Testing the limits of quantum mechanics: motivation, state of play, prospects.
	\textit{Journal of Physics: Condensed Matter} \textbf{14}: R415--R451 (2002)

\bibitem[Lewis(1995)]{Lew95} P. Lewis: 
	GRW and the tails problem.
	\textit{Topoi} \textbf{14}: 23--33 (1995)

\bibitem[Lewis(1997)]{Lew99} P. Lewis:
	Quantum Mechanics, Orthogonality, and Counting.
	\textit{British Journal for the Philosophy of Science} \textbf{48}: 313--328 (1997)

\bibitem[Lewis(2003)]{Lew03} P. Lewis:
	Counting Marbles: A Reply to Critics.
	\textit{British Journal for the Philosophy of Science} \textbf{54}: 165--170 (2003)

\bibitem[Lewis(2005)]{Lew05} P. Lewis: 
	Interpreting Spontaneous Collapse Theories. 
	{\it Studies in History and Philosophy of Modern Physics} \textbf{36}: 165--180 (2005)

\bibitem[Lewis(2006)]{Lew06} P. Lewis:
	GRW: A Case Study in Quantum Ontology.
	\textit{Philosophy Compass} \textbf{1}: 224--244 (2006)

\bibitem[Maudlin(2007)]{Mau06} T. Maudlin:
	Completeness, supervenience and ontology.
	\textit{Journal of Physics A: Mathematical and Theoretical} 
	\textbf{40}: 3151--3171 (2007)

\bibitem[Maudlin(2010)]{Mau10} T. Maudlin: 
	Can the world be only wavefunction?
	Pages 121--143 in S. Saunders, J. Barrett, A. Kent, and D. Wallace (ed.s): 
	{\it Many worlds? Everett, quantum theory, and reality.} 
	Oxford University Press (2010)

\bibitem[Maudlin(2011)]{Mau11} T. Maudlin:
	\textit{Quantum Non-Locality and Relativity: Metaphysical Intimations 
	of Modern Physics}. 3rd edition. 
	Oxford: Blackwell (2011)

\bibitem[Maudlin(2016)]{Mau14} T. Maudlin:
	Local Beables and the Foundations of Physics.
	Pages 317--330 in M.~Bell and S.~Gao (ed.s): 
	\textit{Quantum Nonlocality and Reality -- 50 Years of Bell's Theorem} (2016)
	\url{http://www.ijqf.org/wps/wp-content/uploads/2014/12/Maudlin-Local-Beables.pdf}

\bibitem[McQueen(2015)]{McQu15} K. J. McQueen: 
	Four Tails Problems for Dynamical Collapse Theories.
	\textit{Studies in the History and Philosophy of Modern Physics} \textbf{49}: 10--18 (2015)
	\url{http://arxiv.org/abs/1501.05778}

\bibitem[Monton(2004)]{Mon04} B. Monton: 
	The Problem of Ontology for Spontaneous Collapse Theories.
	\textit{Studies in History and Philosophy of Modern Physics} \textbf{35}: 407--421 (2004)
	\url{http://philsci-archive.pitt.edu/1410/}

\bibitem[Pearle(1989)]{Pe89} P. Pearle: 
	Combining stochastic dynamical state-vector reduction with spontaneous localization. 
	\textit{Physical Review A} \textbf{39}: 2277--2289 (1989)

\bibitem[Pearle(1990)]{Pea90} P. Pearle: 
	Toward a relativistic theory of statevector reduction.
	Pages 193--214 in A. I. Miller (ed.), \textit{Sixty-Two Years of Uncertainty}, 
	volume 226 of \textit{NATO ASI Series B}. 
	New York: Plenum (1990)

\bibitem[Pearle(2009)]{Pea06} P. Pearle: 
	How Stands Collapse II. 
	Pages 257--292 in W. Myrvold and J. Christian (ed.s): 
	\textit{Quantum Reality, Relativistic Causality, and Closing the Epistemic Circle}, 
	The Western Ontario Series in Philosophy of Science Vol. 73. 
	New York: Springer (2009)
	\url{http://arxiv.org/abs/quant-ph/0611212}

\bibitem[Penrose(2000)]{Pen00} R. Penrose: 
	Wavefunction Collapse As a Real Gravitational Effect. 
	Pages 266--282 in A. Fokas, T. W. B. Kibble, A. Grigoriou, B. Zegarlinski (ed.s), 
	\textit{Mathematical Physics 2000}.
	London: Imperial College Press (2000)

\bibitem[Sebens(2015)]{Seb15} C. Sebens:
	Killer Collapse: Empirically Probing the Philosophically Unsatisfactory Region of GRW.
	\textit{Synthese} {\bf 192}: 2599--2615 (2015)
	\url{http://philsci-archive.pitt.edu/11350/}

\bibitem[Shimony(1990)]{Shi} A. Shimony: 
	Desiderata for a Modified Quantum Dynamics.
	Pages 49--59 in A. Fine, M. Forbes and L. Wessels (ed.s), 
	\textit{Proceedings of the 1990 Biennial Meeting of the Philosophy of
	Science Association Vol. 2}. 
	East Leasing: Philosophy of Science Association (1990)

\bibitem[Tumulka(2006a)]{Tum06} R. Tumulka: 
	A relativistic version of the Ghirardi--Rimini--Weber model.
	\textit{Journal of Statistical Physics} \textbf{125}: 821--840 (2006a)
	\url{http://arxiv.org/abs/quant-ph/0406094}

\bibitem[Tumulka(2006b)]{Tum06c} R. Tumulka:
	Collapse and Relativity.  
	Pages 340--352 in A. Bassi, D. D\"urr, T. Weber and N. Zangh{\`{\i}} (eds.), 
  	\textit{Quantum Mechanics: Are there Quantum Jumps? and
  	On the Present Status of Quantum Mechanics}, 
  	AIP Conference Proceedings \textbf{844}. 
	American Institute of Physics (2006b) 
	\url{http://arxiv.org/abs/quant-ph/0602208}

\bibitem[Wallace(2008)]{Wal08} D. Wallace:
	Philosophy of quantum mechanics. 
	Pages 16--98 in D. Rickles (ed.), 
	{\it The Ashgate companion to contemporary philosophy of physics}. 
	Aldershot: Ashgate (2008)
	\url{http://arxiv.org/abs/0712.0149}

\bibitem[Wallace(2014)]{Wal14} D. Wallace:
	Life and death in the tails of the GRW wave function. (2014)
	\url{http://arxiv.org/abs/1407.4746}
	
\bibitem[Weinberg(2012)]{Wein12} S. Weinberg: Collapse of the State Vector.
	\textit{Physical Review A} {\bf 85}: 062116 (2012)
	\url{http://arxiv.org/abs/1109.6462}

\end{thebibliography}
\end{document}